\documentclass[twocolumn]{aastex61}

\usepackage[T1]{fontenc}
\usepackage{apjfonts}
\usepackage[figure,figure*]{hypcap}

\accepted{August 29, 2018}
\submitjournal{ApJ}

\shorttitle{Flux Cancellation in Bipolar ARs}
\shortauthors{Yardley et al.}

\begin{document}

\title{The Role of Flux Cancellation in Eruptions from Bipolar Active Regions}

\author{S. L. Yardley}
\affiliation{University of St Andrews, School of Mathematics and Statistics, North Haugh, St Andrews, Fife KY16 9SS, UK}
\affiliation{University College London, Mullard Space Science Laboratory, Holmbury St. Mary, Dorking, Surrey RH5 6NT, UK}
\author{L. M. Green}
\affiliation{University College London, Mullard Space Science Laboratory, Holmbury St. Mary, Dorking, Surrey RH5 6NT, UK}
\author{L. van Driel-Gesztelyi}
\affiliation{University College London, Mullard Space Science Laboratory, Holmbury St. Mary, Dorking, Surrey RH5 6NT, UK}
\affiliation{Observatoire de Paris, LESIA, UMR 8109 (CNRS), F-92195 Meudon-Principal Cedex, France}
\affiliation{Konkoly Observatory of the Hungarian Academy of Sciences, Budapest, Hungary}
\author{D. R. Williams}
\affiliation{ESA European Space Astronomy Centre, 28692 Villanueva De La Ca\~{n}ada, Madrid, Spain.}
\author{D. H. Mackay}
\affiliation{University of St Andrews, School of Mathematics and Statistics, North Haugh, St Andrews, Fife KY16 9SS, UK}

\begin{abstract}

The physical processes or trigger mechanisms that lead to the eruption of coronal mass ejections (CMEs), the largest eruptive phenomenon in the heliosphere, are still undetermined. Low-altitude magnetic reconnection associated with flux cancellation appears to play an important role in CME occurrence as it can form an eruptive configuration and reduce the magnetic flux that contributes to the overlying, stabilising field. We conduct the first comprehensive study of 20 small bipolar active regions in order to probe the role of flux cancellation as an eruption trigger mechanism. 
We categorise eruptions from the bipolar regions into three types related to location and find that the type of eruption produced depends on the evolutionary stage of the active region. In addition we find that active regions that form eruptive structures by flux cancellation (low-altitude reconnection) had, on average, lower flux cancellation rates than the active region sample as a whole. Therefore, while flux cancellation plays a key role, by itself it is insufficient for the production of an eruption. The results support that although flux cancellation in a sheared arcade may be able to build an eruptive configuration, a successful eruption depends upon the removal of sufficient overlying and stabilising field. Convergence of the bipole polarities also appears to be present in regions that produce an eruption. These findings have important implications for understanding the physical processes that occur on our Sun in relation to CMEs and for space weather forecasting.

\end{abstract}


\section{Introduction}
Coronal mass ejections (CMEs) are the most energetic phemonena in the Solar System, involving around $10^{32}$ ergs of energy in the form of electromagnetic, kinetic, thermal, non-thermal and gravitational potential energy. 
The energy is ultimately derived from the coronal magnetic field, where it is stored in the form of electric currents \citep{Forbes-2000}. However, the exact evolution of the coronal magnetic field, and the physical processes involved in CMEs, are still subjects of study. CMEs are also of interest because they can drive intense geomagnetic storms \citep{Gosling-1993}. These storms are able to create hazardous space weather conditions at Earth, leading to disruptions of our technological systems, and significant socioeconomic impact \citep[for a review see][]{Eastwood-2017}. Understanding the conditions in which CMEs are created is therefore of importance for a physical understanding of our Sun, as well as for space weather forecasting. 

The occurrence of CMEs involves an energy storage-and-release process and their formation is often discussed as having two phases; a trigger and a driver. The trigger refers to the physical process(es) that brings the magnetic field to the point of an eruption, whereas the driver is responsible for the sudden expansion and upward acceleration of the erupting volume. The driving mechanism appears to be limited to either magnetic reconnection taking place in a vertical current sheet below the eruptive structure \citep{Moore-2001} or the Lorentz force acting on a flux rope \citep{Forbes-1991,Torok-2005,Kliem-2006,Mackay-2006a,Mackay-2006b,Kliem-2014}. Possible trigger mechanisms, however, appear to be wide-ranging and include, for example, sunspot rotation, flux emergence and photospheric flows. See \cite{Green-2017} for an overview of CME trigger and driver processes. Our efforts to understand (and forecast) CMEs are severely impeded by a lack of knowledge of the relative importance of these trigger mechanisms.

In this study we focus on another particular CME trigger known as flux cancellation. In the flux cancellation process small-scale opposite polarity magnetic fragments are seen to converge, collide and  disappear along the polarity inversion line (PIL) that separates regions of positive and negative field in the photosphere \citep{Martin-1985}. The disappearance of the two opposite polarity fragments is ultimately the consequence of the fragmentation and dispersion of the magnetic field caused by convective flows and differential rotation. Three scenarios have been proposed to explain the process of flux cancellation (see \cite{Zwaan-1987}): the emergence of a U-loop \citep{vDG-2000,Bernasconi-2002}, the submergence of an $\Omega$-loop below the surface \citep{Harvey-1999,Chae-2004,Yang-2009}, or the result of magnetic reconnection taking place at a low height \citep{vB-1989}. We investigate the third case where flux cancellation due to low-altitude magnetic reconnection is able to gradually transform a sheared arcade field into a flux rope. In this scenario, magnetic reconnection produces two loops: 1) a small loop with a high curvature, which submerges below the photosphere leading to the disappearance of the small bipole; 2) a loop much larger in size-scale that extends into the corona. Ongoing flux cancellation can therefore form a flux rope that is expected to have its underside located in the high plasma-$\beta$ environment of the lower solar atmosphere. During the flux cancellation process an amount of flux equal to that which is cancelled is available to be built into the flux rope. The actual amount of flux that is built into a flux rope depends on the properties of the region, such as the amount of shear and the length of the PIL along which flux cancellation is occurring. The details of this process are discussed in \cite{Green-2011}. 

The flux cancellation process also has a secondary effect in that it reduces the flux in the region that contributes to the field overlying, and stabilising, the flux rope. If enough flux is transformed from the overlying arcade into the flux rope, a force imbalance can occur leading to a catastrophic loss of equilibrium and a CME \citep{Lin-2000, Bobra-2008}. Or, if the active region evolves to a point where the overlying field decreases rapidly enough with height, the flux rope can become torus unstable \citep{Kliem-2006}. In this way, flux cancellation can be viewed as a CME trigger mechanism, which in itself requires a converging flow, in a sheared field, to bring opposite polarity fragments together. Such a scenario for flux rope formation and eruption due to flux cancellation is well supported by simulations \citep{Amari-2003, Aulanier-2010} and observations \citep{Green-2011, Yardley-2016}.

Here we present the first comprehensive study of the eruptive activity in a representative sample of 20 small bipolar active regions (ARs) in order to probe the role of flux cancellation as a CME trigger. We study the evolution of the photospheric magnetic field
to quantify the significance of flux cancellation in building an eruptive magnetic field environment. We investigate at what point in an active region's lifetime eruptions occur.

\section{Data \& Methods}

\subsection{Active Region Selection}

In this study, we focus on eruptions that are produced in bipolar active regions. Bipolar active regions are selected for study due to their low magnetic complexity, which 
minimizes the number of polarity inversion lines (PILs) along which eruptions might originate. Active regions were selected using the following criteria. The regions must have two dominant magnetic polarities with no major mixing of the opposite polarities.  The active regions must be isolated from other active regions so that flux cancellation occurring 
along any external PILs is negligible allowing flux cancellation along the internal PIL to be quantified. 
They must be short-lived regions and form east of central meridian so that their evolution can be tracked across the disk. Finally, the active regions must emerge within 60$^{\circ}$ of central meridian due to the decreasing reliability of the magnetic flux measurements with increasing distance from disk centre.
The above criteria led to the selection of 20 active regions from the HMI era, spanning a time period from March 2012 to November 2015. 

It should be noted that the above criteria necessarily lead to the selection of small active regions, with magnetic flux $\sim 10^{20}$--10$^{21}$~Mx. Eruptions from these regions may produce relatively subtle signatures in extreme ultraviolet data and no observable coronal mass ejection in white light coronagraph data. Due to this, we do not use the term coronal mass ejection (CME) in this work. Rather we refer to eruptions that are identified in extreme ultraviolet data. These eruptions may be successful or may not be fully ejected from the Sun, leading to a failed eruption. With such weak events it can be hard to discriminate these two categories but since we are interested in the role of flux cancellation as a CME trigger, we do not focus on whether each event is failed or fully ejective only whether it was initiated in the first place.


\subsection{Coronal evolution and eruptive activity}

\begin{figure*}[ht]
\centering
\includegraphics[width=\linewidth]{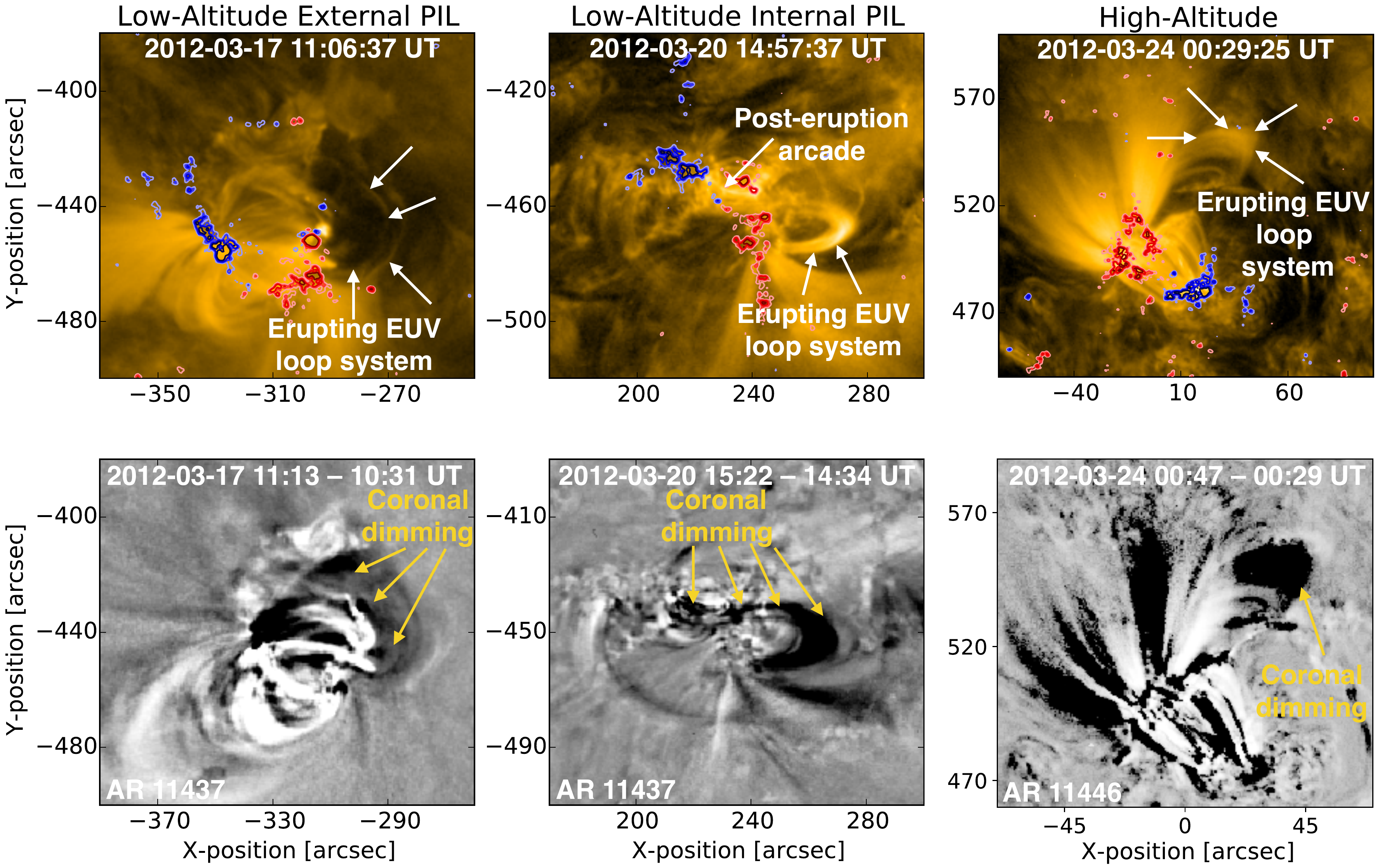}
\caption{Examples of the three eruption categories: low-altitude external PIL, low-altitude internal PIL and high-altitude events. The top row shows {\it SDO}/AIA 171~\AA\ images that have been overlaid with contours the line-of-sight magnetic field from SDO/HMI (shown at a saturation of $\pm$~100~G). The red (blue) contours correspond to positive (negative) magnetic flux, respectively. The coronal signatures observed in each example have been labelled in white along with the NOAA AR number and time of image. The bottom row shows difference images that have been made using the times given in each image. The location of coronal dimmings associated with each eruption are indicated by yellow arrows. An animation of the images in the bottom row is available. From left to right the sequences start at 2012-03-17 10:35:36 UT, 2012-03-20 14:30:36 UT, and 2012-03-24 00:30:36 UT. The sequences end at 2012-03-17 11:30:36 UT, 2012-03-20 15:25:36 UT, and 2012-03-24 01:25:36 UT, respectively. The video duration is 2 seconds.}
\label{fig:fig1}
\end{figure*}

The coronal evolution of each active region is monitored using extreme ultraviolet (EUV) images produced by the Atmospheric Imaging Assembly (AIA; \citealt{Lemen-2012}) instrument on board the {\it Solar Dynamics Observatory (SDO; \citealt{Pesnell-2012})}. The AIA instrument provides full-disk observations with a high spatial resolution and temporal cadence of 1.5" and 12~s, respectively, for three UV--visible and seven EUV bandpasses. In this study we focus on using 171~\AA\ and 193~\AA\ to analyse the coronal evolution of each active region. The 171~\AA\ passband is dominated by plasma emission at a temperature of around 0.6~MK whereas, the temperature response of the 193~\AA\ has two temperature peaks at approximately 1.2 and 20~MK. 

Each active region is analysed in order to identify the time and location of eruptions that are produced. Coronal signatures used to indicate the occurrence of an eruption include at least two of the following: the eruption of a filament or EUV loop system, the rapid disappearance of coronal loops and the formation of a post-eruption arcade (flare arcade), flare ribbons and coronal dimmings. 


Each eruption is then placed into one of three categories; the aim being to identify which eruptive structures form at a low-altitude, and can therefore be studied in the context of flux cancellation taking place in the active region, and which are formed by other processes and/or altitudes. The three categories are given the names: internal PIL events (for the eruption of a low-altitude structure from along the active region's internal PIL); external PIL events (for the eruption of a low-altitude structure along a PIL at the periphery of the active region), high-altitude events (for the eruption of a structure from a high-altitude in the corona and presumably not associated with flux cancellation during the time period studied).


One or more of the following criteria must be met for an eruption to be classified as an internal or external PIL (low-altitude) event:
\begin{itemize}
\item The low-lying core field of the active region must be opened and reconfigured as new post-eruption (flare) loops form.
\item Any flare ribbons that form must, in the first instance, be immediately next to and run along the PIL.
\item Any dimming regions that form must, in the first instance, be immediately next to the PIL.
\end{itemize}

One or more of the following criteria must be met for an eruption to be classified as a high-altitude event:
\begin{itemize}
\item The low-lying core field of the active region must not be involved or modified.
\item Any flare ribbons that form must be well separated from the PIL.
\item Any post-eruption (flare) loops that form must be located above the active region core field.
\item Any dimming regions that form must, in the first instance, be remote from the PIL.
\end{itemize}

One example from each eruption category is shown in Figure~\ref{fig:fig1} and online in Supplementary Movie 1.

\subsection{Magnetic Flux Evolution} \label{sec:mfe}

The photospheric field evolution of each active region is analysed using magnetograms obtained by the Helioseismic and Magnetic Imager (HMI; \citealt{Schou-2012}) on board {\it SDO}. The magnetograms used in this study provide information on the line-of-sight component of the magnetic field and are from the 720~s series (hmi.M\_720s) recorded by the vector camera. This camera has a pixel size of 0.5" and a noise level of 10~G. 
The magnetic flux evolution of each active region was calculated by implementing the Solar Tracking of the Evolution of Photospheric Flux (STEF) algorithm \citep{Yardley-2016} on the line-of-sight magnetograms. The cadence of the magnetograms used is 96 minutes. 

Active regions are manually identified in the full-disk magnetograms and a field of view is assigned as a rectangular box centered on the active region. The radial component of the magnetic field $B_{R}$ is estimated for each pixel in the series of full-disk magnetograms by applying a cosine correction to the longitudinal magnetic field $B_{LOS}$ using Heliocentric Earth Equatorial Coordinates (HEEQ):
\begin{equation}
B_{R} = \frac{B_{LOS}}{\cos{\theta} \cos{\phi}},
\end{equation}
where $\theta$ and $\phi$ are the helioprojective westward and northward angles, respectively. These angles can be expressed in terms of heliocentric westward and northward coordinates x, y
\begin{eqnarray}
\theta & = & \arcsin{\frac{x}{R \cos{\phi}}}, \\
\phi & = & \arcsin{\frac{y}{R}},
\end{eqnarray}
where $R$ is the radius of the Sun with respect to the observer. The magnetogram containing the radialised field values is then de-rotated to the central meridian passage time of the active region to correct for projection effects using a routine that has been developed in SunPy \citep{SunPy-2015}.

\begin{figure*}[ht]
\centering
\includegraphics[width=\linewidth]{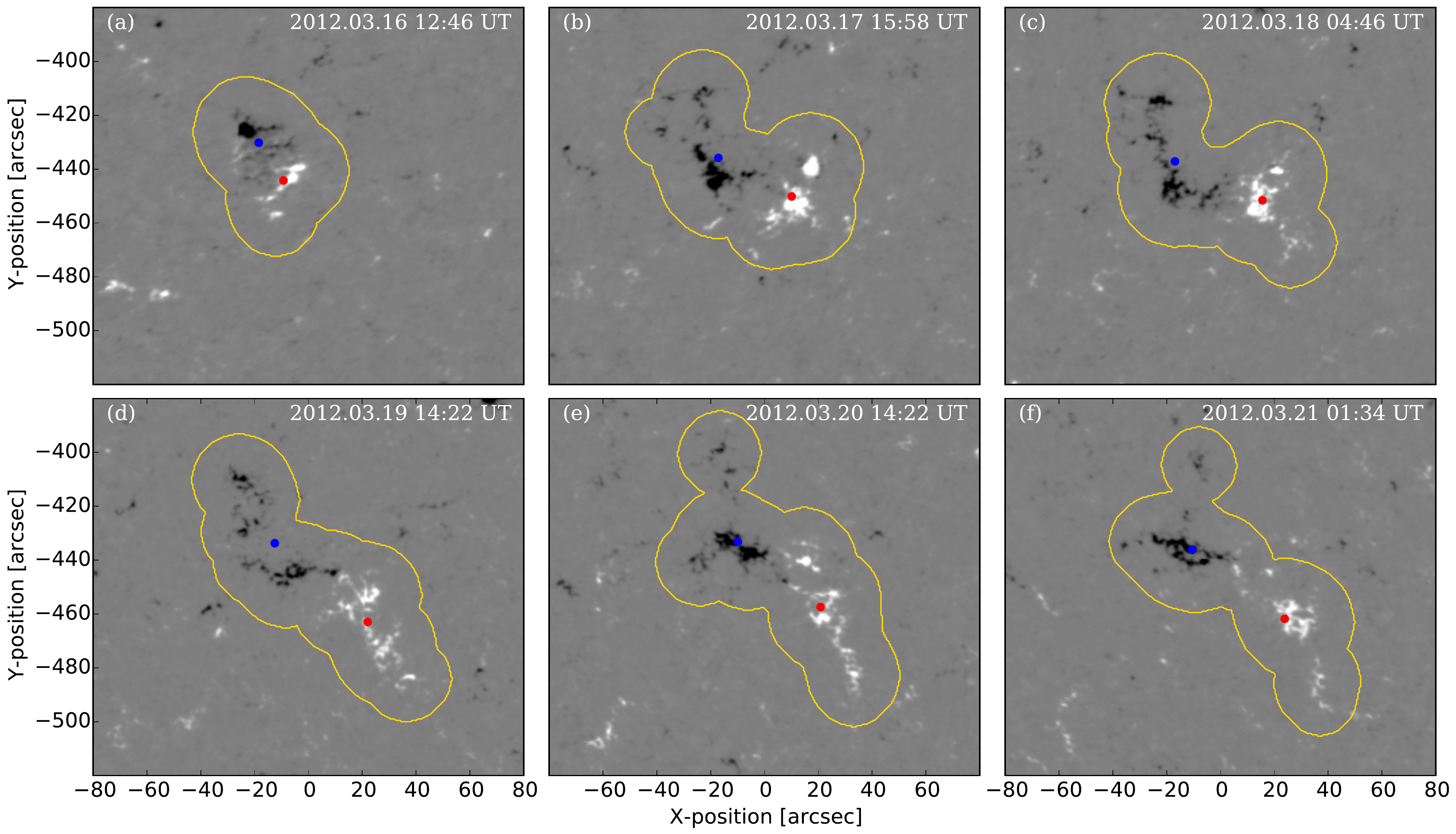}
\caption{Evolution of the line-of-sight magnetic field of NOAA AR 11437 observed using \textit{SDO}/HMI magnetograms. The magnetic field directed towards (away from) the observer is shown in white (black). The active region shows a typical evolution whereby the polarities emerge and separate. The magnetic flux then disperses over time due to the action of photospheric flows and is reprocessed by granular and supergranular convection \citep{Dacie-2016}. The yellow contour represents the region in which the positive, negative and unsigned flux is calculated. The red (blue) points represent the positive (negative) flux-weighted central coordinates. 
}
\label{fig:fig2}
\end{figure*}

The flux-weighted central coordinates of the selected field of view are calculated for each time step making it possible to track the active region such that it always remains in the field of view. The pixels that make up the active region are then selected as follows. First a Gaussian filter is applied to smooth the data with a standard deviation (width) of 7 pixel units. The weighted average of the magnetic flux density of the neighbouring pixels must exceed a threshold of 40 G. This threshold is set manually and has been tested to give the best results. The largest regions of magnetic flux that form at least 60\% of the selected pixels are identified and retained whereas, the smaller features at large distances are disregarded. This is to ensure that quiet sun magnetic features that are not part of the AR are removed. It is still possible that small-scale magnetic features can enter or exit the boundary, introducing a contribution to or reduction of the magnetic flux measurement. These fluctuations are usually small and an error on the flux measurement is estimated by measuring the magnetic flux of small flux fragments that move into or out of the active region area. This error estimation varies in time as the active region evolves. If the automatic detection fails to successfully select the active region flux at any time step a function is used that allows the user to manually select contours of magnetic flux for the flux calculation. 
Finally, 
a dilation is applied so that pixels within approximately 5" of those selected are also included within the active region area selection. Figure~\ref{fig:fig2} shows the line of sight photospheric magnetic field evolution of AR 11437 and the active region area (shown by the yellow contour) identified by the STEF algorithm. Magnetic flux is measured in this area.

The emergence of small bipoles in or close to an active region, or quiet sun fragments that cancel with flux at the periphery of the active region, can affect the measurement of magnetic flux. In these cases it is not possible to accurately identify and measure the flux cancellation at the internal PIL. 
To take this into account, if possible, the flux cancellation is not calculated during time periods when external flux cancellation or flux emergence is occurring. 

\subsection{Separation Distance of AR Polarities}

To determine whether there is an overall convergence of the two polarities of the bipole, the separation of the positive and negative polarities is calculated. The separation distance is quantified by computing the separation of the flux-weighted central coordinates of the positive and negative bipoles. The flux-weighted central coordinates $f$ are computed as follows
\begin{equation}
f = \frac{\sum_{i}B_{i}x_{i}}{\sum_{i} B_{i}},
\end{equation}
where $B_{i}$ and $x_{i}$ are the magnetic flux density and coordinate corresponding to pixel ${i}$ of the active region. The flux-weighted central coordinates are computed in both $x$ and $y$. The separation distance $d$ of the polarities is then calculated by
\begin{equation}
d = \sqrt{(x_{1}-x_{2})^{2} + (y_{1} - y_{2})^{2}},
\end{equation}
where ($x_{1}$,$y_{1}$) and ($x_{2}$,$y_{2}$) are the flux-weighted central coordinates of the positive and negative polarities, respectively.

\section{Results}
\subsection{Active Region Eruptions}

Of the 20 active regions studied, 13 produced at least one eruption. These 13 active regions produced a total of 24 eruptions during the time period studied. Eight out of these 13 active regions produced low-altitude events that originated from either the internal or external PIL of the active region. The remaining five of these 13 active regions produced high-altitude events.
Table~\ref{tab:table1} gives the timings of the different types of event as determined by the eruption and rapid disappearance of EUV coronal loops and also the coronal signatures observed during eruption. The majority of the eruptions (22/24) that occurred exhibited coronal dimmings, which suggests that these events may have been successful CMEs. However, two of the eruptions are not associated with coronal dimmings and are therefore assumed to be confined or failed eruptions. 

\begin{figure*}[ht]
\epsscale{1.1}
\centering
\includegraphics[width=\linewidth]{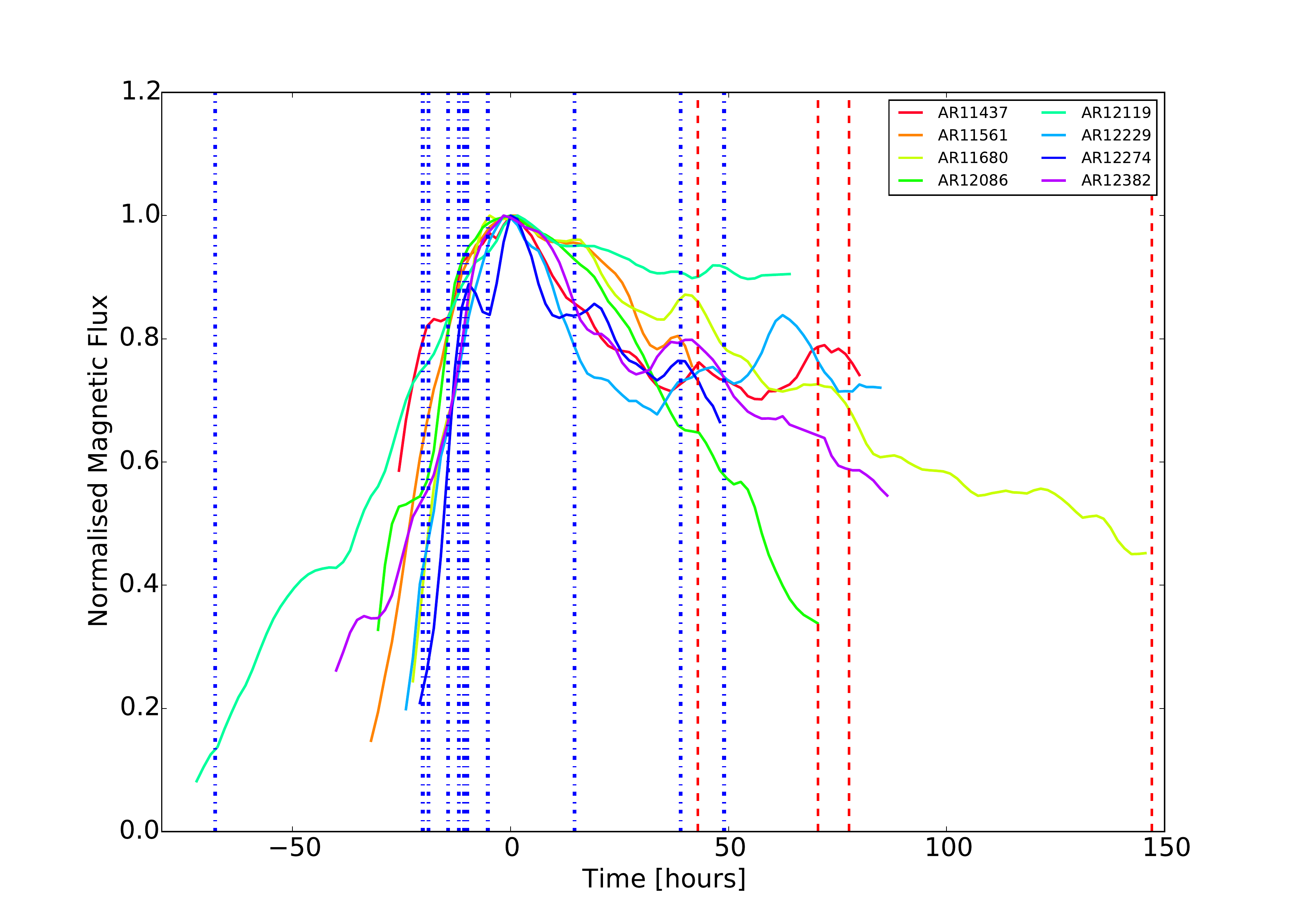}
\caption{Normalised magnetic flux evolution as a function of time for the active regions that produce internal and/or external PIL eruptions. The magnetic flux value and time have been normalised with respect to the peak flux of each active region. The red dashed and blue dot-dashed lines represent the timings of the internal and external PIL eruptions, respectively.}
\label{fig:fig3}
\epsscale{1}
\end{figure*}

It is notable that, during the time period studied, the active regions that produced eruptions produced \textit{either} internal and/or external PIL events (which originate at a low-altitude) \textit{or} high-altitude events. No active regions produced both low-altitude and high-altitude eruptions.
Overall, eruptions occurred in both the emergence and decay phases of the active regions, however the category of event produced depended on the region's evolutionary stage. Table~\ref{tab:CME_AR_phase} and Figure~\ref{fig:fig3} show that there is a tendency for external PIL events to occur during an active region's emergence phase and for internal PIL events to form during the decay phase. 
In contrast to this, the eruption of a high-altitude structure occurs roughly evenly across the emergence and decay phases. The high-altitude events make up only 38\% of the eruptions originating in the active regions.

\subsection{Flux Cancellation} \label{sec:canc_res}

The flux cancellation rate and total flux cancelled in each active region is determined from the reduction in the total unsigned magnetic flux with time (see Figure~\ref{fig:fig4} (a)). That is, during the decay phase of each active region. Therefore, the results presented here only refer to the decay phase of the active regions.

Flux cancellation, at either an internal or external PIL, was observed to occur in all but one active region (AR 11867).  
Table \ref{tab:table3} summarises the flux cancellation rates (column 7) and total amount cancelled (column 8). 
The average flux cancellation rate for all 20 active regions (including AR 11867 that exhibited no cancellation) is 0.84$\times$10$^{19}$~Mx~h$^{-1}$. This compares to an average flux cancellation rate of 0.68$\times$10$^{19}$~Mx~h$^{-1}$ for regions that produce internal and external PIL events, 1.10$\times$10$^{19}$~Mx~h$^{-1}$ for regions that produce high-altitude events, and 0.83$\times$10$^{19}$~Mx~h$^{-1}$ for regions that produce no eruptions at all.
Therefore, active regions that do not produce eruptions have a flux cancellation rate close to the average value across all regions, internal and external PIL event active regions have a flux cancellation rate less than average, and high-altitude event active regions have a flux cancellation rate above the average value.

Although flux cancellation can play an important role in the creation of a sheared or twisted pre-eruptive structure \citep{vB-1989}, what is important for the occurrence of an eruption is the force balance between this structure and the overlying arcade field. To investigate this, the {\it total} amount of flux cancelled must be considered, as this quantity represents the amount of flux that could have been built into a flux rope. This value can then be compared to the amount of flux that remains in the active region as field overlying the rope. Therefore, we now pay particular attention to the four active regions that produce eruptions from their internal PIL, in order to probe how much flux may have been built into the pre-eruptive structure in relation to that remaining as active region arcade field. 
We compare these active regions to those that produce high-altitude eruptions. These two groups exhibit a similar amount of total flux cancelled at the internal PIL over the time period studied, despite their differing flux cancellation rates. They therefore
provide an interesting dataset to compare and contrast regions where flux cancellation apparently acts as an eruption trigger (through the creation and eruption of a low-altitude structure) and a group of regions where it does not.

The flux cancellation measurements for the internal PIL event regions and high-altitude event regions are shown in Table~\ref{tab:flux_canc}. The average total flux cancelled for the internal PIL event regions and for high-altitude event regions is 4.8$\times$10$^{20}$~Mx and 5.3$\times$10$^{20}$~Mx, respectively.
The total amount of flux cancelled as a quantity and a percentage of the active region's peak flux value can be seen in the third and fourth columns of Table~\ref{tab:flux_canc}. The percentage values range from 28\% to 49\% for the low-altitude internal PIL events, which is higher than the average value of 24\% of high-altitude event active regions.

Now we compare the amount of total flux cancelled (available to be built into the flux rope) to that left in the active region as field overlying (and stabilising) magnetic field for internal PIL eruptions and high-altitude eruptions. For AR 11561, a corrected flux cancellation value was used to account for the fact that the flux cancellation measurement could not be made during the entire time that cancellation was observed to occur. This is due to the emergence of a bipole to the south of the active region, which cannot be removed from the magnetic flux measurement and therefore masks the flux cancellation taking place at the internal PIL. 
We find that the ratio of flux cancelled compared to that which remains as overlying field
for active regions that produce internal PIL events (at the time of eruption) lies between 1:0.03 and 1:1.57 (see column 5 of Table~\ref{tab:flux_canc}). Here we note that the ratio of 1:0.03 for AR 11437 suggests that practically no overlying field remains, indicating that the assumption that flux cancellation injects an equal amount of flux into the rope may not apply. This compares to ratios between 1:0.94 and 1:3.42 for regions that produce high-altitude eruptions. Therefore, on average, high-altitude event regions had a relatively high value of flux overlying the PIL along which flux cancellation was occurring.

\subsection{Configuration and Motion of Active Region Polarities}


The orientation of the bipole with respect to the polarity inversion line provides information on the level of non-potentiality (or shear) of the magnetic field. A sheared field is an essential component of the flux rope formation model that is relevant to this study \citep{vB-1989}. We use the observational proxy of sheared loops to determine the non-potentiality of the magnetic field of the bipole. The shear angle was measured for each active region at the time of peak magnetic flux (see column 11 of Table \ref{tab:table3}). The magnetic shear angle is defined as the angle between the normal to the line that connects the flux-weighted central coordinates and the PIL, where the clockwise direction corresponds to a positive shear angle. Nineteen out of the 20 active regions studied showed flux cancellation and the amount of shear in these 19 regions ranges from 1$^{\circ}$ to 40$^{\circ}$.
We found that, on average, the shear angle of the active regions that produced internal PIL events is significantly higher than the regions that produced high-altitude events. On average, the shear angle for regions that produced internal PIL events and those that produced high-altitude events was found to be 28$^{\circ}$ and 14$^{\circ}$, respectively. 

\begin{figure*}[ht]
\centering
\includegraphics[width=0.8\linewidth]{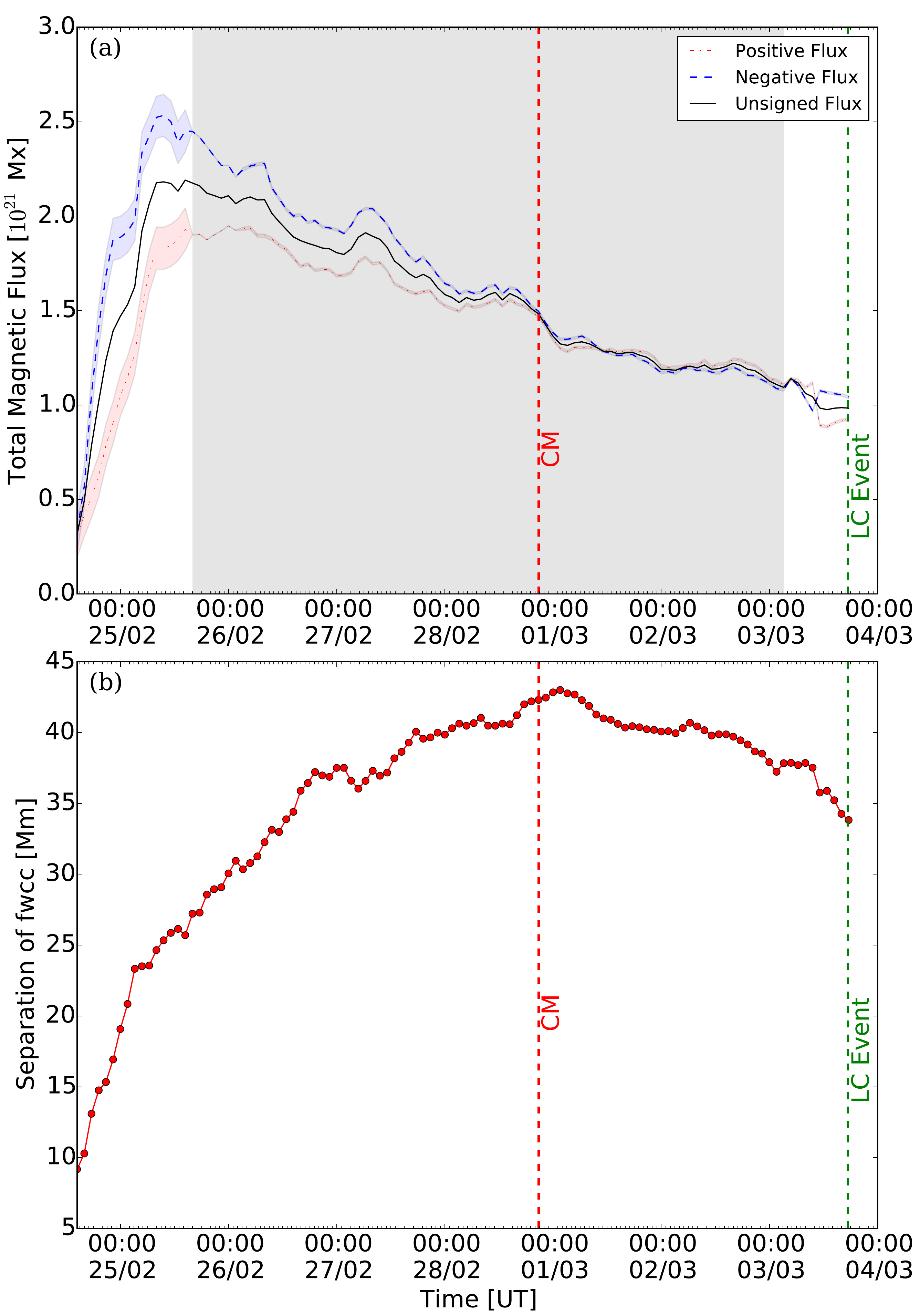}
\caption{ (a) The evolution of magnetic flux in AR 11680. The total positive (red), negative (blue) and unsigned (black) magnetic flux is plotted with the red (blue) shaded regions indicating errors in positive (negative) flux measurements. Error calculation is described in Section \ref{sec:mfe}. The grey shaded region represents the time period over which the flux cancellation was calculated. The red dashed line indicates when the active region crosses central meridian (CM) and the green dashed line the time of an eruption originating from the active region's internal PIL at low-altitude (LC event). (b) The separation of flux-weighted central coordinates for AR 11680. The markings CM and LC E have the same meaning as in (a).}
\label{fig:fig4}
\end{figure*}

Motion toward the polarity inversion line is also an essential component of the model of \citet{vB-1989} as small fragments of the bipole's magnetic field, which break away from the main concentrations, converge, collide and cancel. In addition, an overall convergence of the polarities leads to an inflation of the field which may in turn affect the stability of any flux rope that has formed. Therefore, bipole convergence may also be investigated as a stability proxy. The separation of the active region polarities over time was calculated using the flux weighted central coordinates of the negative and positive flux regions (see Figure~\ref{fig:fig4} (b)). Overall, 75\% of the active regions that produce internal PIL events showed a combination of bipole convergence, shear and flux cancellation. This is much higher than the sub-set of active regions that do not produce internal PIL events (34\%). This sub-set includes regions that produce no eruptions, external PIL eruptions and high-altitude eruptions.


\section{Discussion}


In this study, we investigate the role of flux cancellation as an eruption trigger in a survey of 20 isolated and small bipolar active regions.
Nineteen active regions exhibit flux cancellation, the amount of which was quantified from the reduction in the total unsigned magnetic flux with time. This approach is based on the assumption that flux cancellation is the only process by which active region flux is removed from the photosphere on the timescale of a few days. Other mechanisms of removing flux from an active region include the fragmentation and advection of fragments across larger and larger areas by plasma flows. However, these flux fragments are captured in our method of flux measurement. In addition, Ohmic diffusion will cause flux to diffuse through the photosphere, due to the finite electrical resistance of the plasma. This diffusion process occurs on a timescale $t_{D}$, which is given by $t_{D}=L^{2}/\eta$, where L is the length-scale and $\eta$ is the magnetic diffusivity. However, for a sunspot of length-scale 3000~km and using a value of Ohmic diffusion of $\eta=300~$m$^{2}$~s$^{-1}$ gives a large diffusion timescale of the order 1000 years. 
By definition, flux cancellation as determined by our method can only be calculated during an active region's decay phase, when no new significant flux is emerging into the region and the overall flux value is reducing. 

We also take into account that even though HMI produces high-quality data products, there are uncertainties and systematic errors present in the line-of-sight magnetic flux measurements. The selection criteria imposed when choosing active regions suitable for the study included that the regions had to emerge between $\pm$60$^{\circ}$. This was to avoid the appearance of symmetric peaks, centred around $\sim$60$^{\circ}$ with respect to central meridian due to the sensitivity of the HMI instrument being dependent upon longitude \citep{Hoeksema-2014, Couvidat-2016}. The increase in flux is caused by the increase in value by a few tens of percent of low to moderate flux densities between 250 and 750~G. However, this effect is still present before the active region reaches 60$^{\circ}$. A recent study by \cite{Falconer-2016} has used a sample of 272 large active regions to reduce the net projection error 
in parameters measured from deprojected {\it SDO}/HMI vector magnetograms. They remove the average projection error in an active region's total magnetic flux by assuming that the centre-to-limb curve of the average of the absolute values of magnetic flux of a large number of active regions, which is normalised to the value at central meridian for each AR, gives the average fractional projection error at each radial distance from disk centre. In this study we have not followed the method of \cite{Falconer-2016} as we have only analysed flux cancellation that occurs between $\sim  \pm$45$^{\circ}$.
There are also sinusoidal oscillations with periods of 12 and 24 hours in the evolution of total magnetic flux. This time-varying systematic error is mainly caused by the geosynchronous orbit of the {\textit SDO} spacecraft \citep{Hoeksema-2014}.

Since we have selected isolated active regions and studied the flux cancellation that occurred during their decay phase, we are able to probe the characteristics of the active regions that produced eruptions from a low-altitude along their internal PILs and those that did not.
Here we single out four active regions that produce internal PIL eruptions (ARs 11437, 11561, 11680 and 12382) and five active regions that produce only high-altitude eruptions (ARs 11446, 11808, 11881, 11886) and analyse their decay phase. In Section~\ref{sec:canc_res} we described that both groups have roughly the same amount of total flux cancelled although the active regions that produce high-altitude eruptions have, on average, a higher flux cancellation rate. These two groups of active region have a similar photospheric field evolution but markedly different outcomes in the evolution of the coronal field. These results lead to two questions. Why do active regions with a higher flux cancellation rate during their decay phase produce no eruptions from their internal PILs as the \cite{vB-1989} flux rope model might suggest? What are the distinguishing features between these two groups of active regions?

These questions can be addressed by considering the ratio of cancelled flux that is available to be built into the flux rope, versus the remaining flux in the overlying arcade. When more active region flux is cancelled and built into a flux rope, there is less overlying field remaining in the active region to stabilise the structure. Previous observational flux cancellation studies have found a ratio of flux contained in the rope compared to the flux remaining in the overlying arcade of 1:0.65 \citep{Green-2011} and 1:0.9 \citep{Yardley-2016}. Whereas, studies from a modelling perspective have yielded values between 1:1.5 and 1:1.9 \citep{Bobra-2008, Savcheva-2009, Savcheva-2012}. In this case we found that the ratio of flux cancelled (i.e. the flux available to be built into the rope) compared to that in the overlying field at the time of the internal PIL event is: 1:1.29, 1:1.57, 1:0.03, 1:0.32 for ARs 11437, 11561, 11680 and 12382, respectively. We note that whilst for ARs 11437 and 11561 the ratio is very similar to previous results, for active regions 11680 and 12382 the flux contained in the overlying arcade is very small. This suggests that the assumption that flux cancellation injects an equal amount of flux into the rope as that cancelled may not fully apply here. 

This is due to the fact that the total flux cancelled is equal to the amount of flux that is {\it available} to be built into the flux rope, and therefore represents an upper limit on the flux that has been built into the rope. However, the actual amount that builds into the rope is dependent upon the shear of the arcade and the length of the active section of the PIL where flux cancellation is taking place \citep{Green-2011}. Both of these parameters can vary during the lifetime of an active region. An increase in the active section of the PIL and the shear of the arcade field can increase the chances of a loop being involved in a flux cancellation event at both of its ends. When this is the case, flux is cancelled without contributing to the amount of flux in the rope. AR 11680 exhibits a strong increase in the length of the active section of the PIL and AR 12382 shows a large increase in shear between the positive and negative polarities meaning that the amount of flux being built into rope may be overestimated.

Active regions that produce high-altitude events have a larger proportion of flux remaining in the overlying arcade compared to regions that produce internal PIL events. AR 11881 is an outlier in terms of the ratio for the high-altitude regions as it has a value of 1:0.94, which is within the range that produce internal-PIL eruptions. However, when analysing the AIA data in the time period following the end of our flux cancellation measurement we observe an internal PIL eruption that occurs on 2013 October 31 at around 01:50~UT. This is just over a day after our flux cancellation measurements ceased because the magnetic flux evolution could no longer be followed.
There were no internal PIL events observed following the end of the flux cancellation measurement for the remaining active regions that produced high-altitude eruptions.
Our results also show that the average shear angle of the active regions that produce internal PIL events is, on average, higher than that of the other event categories. These results suggest that flux cancellation within a sheared arcade may build a potentially eruptive configuration but that a successful eruption depends on the removal of sufficient overlying and stabilising field.

In a recent study by \citet{Sterling-2017} the evolution of a series of coronal jets that occurred at the periphery of the leading sunspot of AR 12259 were analysed. They found that seven active region jets occurred during strong flux cancellation calculating an average flux cancellation rate of 1.5 $\times$ 10$^{19}$~Mx~h$^{-1}$ with an average of $\sim$5 $\times$ 10$^{18}$~Mx cancelled prior to each episode. The flux cancellation rates for the active region jets were found to be higher than the active regions in this study. This is not that surprising considering that the photospheric evolution of the jet-productive area is on the same size-scale as the active regions and the area is followed for a period of hours rather than days. On average, the total flux cancelled in the active regions in this study was found to be 2 orders of magnitude larger than for the active region jets.

In this study we have focussed on the flux cancellation scenario of \cite{vB-1989} and the role it plays in the productivity of eruptions in small and isolated bipolar ARs. This required an analysis of the relationship between flux cancellation, the evolution of the coronal magnetic field and eruption onset. We conclude that flux cancellation can be considered as a CME trigger if sufficient stabilising field is removed from above the sheared core field.
Other studies have investigated which non-potentiality parameters are strong indicators that a CME will occur. For example, \citet{Bobra-2016} used features derived from {\it SDO}/HMI vector magnetograms to deduce whether active regions that produce M1 class flares or above will also produce a CME. They determined which features distinguish flaring active regions that produce CMEs from those that do not. The study found that the highest-performing features, which characterise the non-potentiality of the magnetic field, are the mean horizontal gradient of the magnetic field and the twist parameter. A study by \citet{Tiwari-2015} found that active regions with a larger non-potentiality and total magnetic flux can produce both fast and slow CMEs, whereas smaller active regions with a more potential configuration can only produce slower CMEs. One key factor that plays a key role in CME productivity is the configuration of the overlying field \citep{Torok-2005}. The gradient of the overlying field with height, for the active regions in our study, will be investigated in the future using non-linear force-free modelling.

\section{Summary}

In this study, we analysed a sample of 20 bipolar active regions over several days starting at the time of emergence, in order to investigate the importance of flux cancellation as a CME trigger. Following the model of \cite{vB-1989}, flux cancellation is the result of magnetic reconnection that is driven by the convergence and collision of loop footpoints in a sheared arcade. 
This process is able to build an eruptive structure, cut the tethers of the overlying field and hence act as a CME trigger. 

Flux cancellation was observed in all active regions with the exception of AR 11867, which remained in its emergence phase during the time period studied. In total, 24 eruptions were produced in 13 active regions. These eruptions were categorised into three types: low-altitude eruptions from the internal PIL (internal PIL events), low-altitude eruptions from an external PIL (external PIL events) and eruptions originating from high in the corona (high-altitude events). We found that the category of eruption produced is related to the evolutionary stage of the bipolar active region. For example, the majority of external PIL events occurred during the active region's emergence phase, when the growing bipole pushes into the surrounding, pre-existing magnetic field. This interaction can drive reconnection and flux cancellation, which may build an eruptive structure at the edge of the active region. The three external PIL events that occurred during the decay phase  of their host active region appear to be related to the emergence of a small bipole close to the active region periphery. This forms an external PIL where cancellation can occur between opposite polarities. In contrast, internal PIL events only occurred during the bipole's decay phase when the bipole polarities have stopped separating and fragmented active region flux is being brought to the internal PIL via convective flows, driving reconnection and flux cancellation. High-altitude events occurred during both the emergence and decay phase of the active regions and could be the result of the destabilisation of a pre-existing structure or the formation of a high-altitude structure formed during the evolution of the active region. We do not carry out an in-depth study of this category of eruption here, rather we leave this to a subsequent study. In summary, we found that no active regions produced both low- and high-altitude eruptions during the time they were studied and that low-altitude eruptions originating from external PILs were more common than those from internal PILs.

Internal PIL and external PIL eruption-productive regions had, on average, lower flux cancellation rates than the active regions that produced high-altitude eruptions and regions that produced no eruptions at all. However, the regions that produced low- or high-altitude eruptions exhibited a similar amount of total flux cancelled, despite the difference in cancellation rates. Therefore, a high rate of flux cancellation and associated reconnection is not alone a sufficient condition for eruption. 
For the four active regions that produced low-altitude events originating along the internal PIL we found that on average 36\% of the peak active region flux had cancelled prior to eruption. This is consistent with percentages found in previous studies \citep{Green-2011, Baker-2012, Yardley-2016} and is 12\% higher than the average value across non-internal PIL event active regions.

A secondary effect of the flux cancellation process is the reduction of active region flux that contributes to the magnetic field overlying and stabilising the sheared core structure (that might contain a flux rope). If a flux rope has formed, and sufficient magnetic flux is transferred from the overlying arcade to this rope, an eruption may be produced. The quantity of flux cancelled, which is equal to the amount available to be built into the flux rope, compared to the flux of the remaining overlying field was found to be in the range 1:0.03 -- 1:1.57 and 1:2.26 -- 1:3.42 for active regions that produced low-altitude internal PIL and high-altitude events, respectively. The ratio for AR 11881 has been omitted from consideration here because this region produced a low-altitude eruption originating from the internal PIL just over a day after the end of the flux cancellation measurement. Therefore, we find that a successful eruption originating from a low-altitude at the internal PIL depends upon the removal of a significant amount of the overlying field, which otherwise acts to stabilise the flux rope.

The non-potentiality (or shear) of the arcade field is a key aspect of the \cite{vB-1989} flux rope model. The shear angle measured at the time of the peak active region flux is, on average, 14$^{\circ}$ higher for the regions that produced low-altitude internal PIL events compared to the regions that produced eruptions at high-altitude.
An overall convergence of the bipole is important for the gradient of the coronal field that acts to stabilise any flux rope formed. We find that 75\% of the active regions that produced low-altitude eruptions along their internal PIL showed a combination of bipole convergence, shear and flux cancellation. Only 34\% of the active regions that do not produce internal PIL eruptions show this combination.

In summary, we have conducted the first extensive study of eruptive activity in a sample of 20 small bipolar active regions taken from the {\it SDO}/HMI era in order to probe the role of flux cancellation as a CME trigger. The results of this study lead to the conclusion that although flux cancellation plays a key role it is not alone sufficient in the production of low-altitude eruptions. A combination of ongoing flux cancellation in a sheared arcade, which is consistent with the flux rope model of \citet{vB-1989}, can build a pre-eruptive configuration but a successful eruption depends upon the removal of sufficient overlying field that would otherwise stabilise the underlying flux rope. In this study the cancellation of more than $\sim$30\% of the peak active region flux value appeared to be sufficient. In addition the eruptions appear to be aided by the convergence of the bipole polarities.

\pagebreak

\begin{table*}[ht]
\centering
\begin{tabular}{|c|c|c|c|c|}
\hline
NOAA & Internal & External & High-Altitude & Coronal \\
AR &  PIL Event Timings (UT) & PIL Event Timings (UT) & Event Timings (UT) & Signatures \\
\hline

11437 &	- & 2012-03-17 05:14  & - & A, D, L, R \\
      & - & 2012-03-17 10:53  & - & A, D, L, R \\     
      & 2012-03-20 14:46 & - & - & A, D, L \\
11446 & - & - & 2012-03-24 00:42 & A, D, L \\
11561 & 2012-09-01 23:37 & - & - & A, D, L, R\\
11680 & 2013-03-03 17:27* & - & - & A, D, F, R \\
11808 & - & - & 2013-07-30 04:04 & A, D, L, R \\
      & - & - & 2013-07-31 15:10 & A, D, L \\
11881 & - & - & 2013-10-24 08:10 (09:12) & A, D, L \\
      & - & - & 2013-10-27 19:45 & A, D, L \\
      &-  & - & 2013-10-29 02:58 & A, D \\
11886 & - & - & 2013-10-29 12:57 & D \\
12086 & - & 2014-06-10 14:49 & - & D, L, R \\
12119 & - & 2014-07-18 10:40 & - & A, D, R \\
      & - & 2014-07-22 21:02 & - & D, L, R  \\
      & - & 2014-07-23 07:30 (08:12) & - & A, D, L, R \\
12229 & - & 2014-12-05 03:46  & - & A, D, R\\
      & - & 2014-12-05 08:12  & - & A, D, R \\
      & - & 2014-12-05 10:39*  & - & R (C) \\   
      & - & 2014-12-05 12:35  & - & D, R \\  
12274 & - & 2015-01-25 20:00 & - & L, R (C) \\
12336 & - & - & 2015-05-05 01:29 & A, D, L \\
      & - & - & 2015-05-05 09:24 & A, D, L \\
12382 & 2015-07-09 02:29 & - & - & A, D, L, R \\
\hline
\end{tabular}
\caption{This table details the eruptions that occurred during the emergence and decay phases of each active region. See Table~\ref{tab:table3} for related active region and flux cancellation information. Column 1 indicates NOAA active region number, columns 2, 3 and 4 indicate the timings of internal PIL events, external PIL events and high-altitude events respectively. Column 4 details the onset time of each eruption as determined by the rapid expansion and eruption of EUV loops. Column 5 gives the coronal signatures observed during an eruption including the eruption of a filament (F), the eruption of an EUV loop system or rapid disappearance of coronal loops (L), post-eruption (flare) arcade or loops (A), flare ribbons (R) and coronal dimming(s) (D). Eruptions that showed no clear coronal dimmings are indicated by (C) in column 5. These could be confined or failed eruptions. *The timings of the eruption onset for one internal PIL eruption originating from AR 11680 and one external PIL eruption from AR 12229 were determined by the onset of a filament eruption. One eruption that originated from AR 11881 and one from AR 12119 could also be observed in LASCO/C2. The timings for the LASCO/C2 observations are given in brackets after the timings of eruption onset. \label{tab:table1}}
\end{table*} 

\begin{table}[h!]
\centering
\begin{tabular}{|c|c|c|}
\hline
              & Emergence & Decay  \\
              & phase     & phase  \\
\hline
External PIL  &  8        & 3      \\
Internal PIL  &  0        & 4      \\
High-altitude &  5        & 4      \\
\hline
\end{tabular}
\caption{\label{tab:CME_AR_phase} 
This table details when internal PIL, external PIL and high-altitude eruptions occur in relation to the evolutionary phase of their source active region.}
\end{table}

\begin{table*}[htbp!]
\begin{rotatetable*}
\scriptsize
\setlength{\tabcolsep}{2pt}
\begin{tabular}{*{12}{|c}|}
\hline
NOAA & Heliographic & Emergence & Peak Flux & Flux Cancellation & Flux Cancellation & Flux Cancellation & Total Flux & Percentage of & FR vs. & Shear & AR \\
AR & Coordinates & Time & Time & Start Time & End Time & Rate & Cancelled & Peak AR Flux & Overlying  & Angle at & Bipole  \\
No. & ($\theta$,$\phi$) & (UT) & (UT) & (UT) & (UT) & (10$^{19}$~Mx~h$^{-1}$) & (10$^{21}$~Mx) & Cancelled (\%) & AR Ratio & Peak Flux ({}$^{\circ}$) & Convergence \\
\hline
11437 & S29 E33 & 2012-03-16 12:46 & 2012-03-17 15:58 & 2012-03-17 15:58 & 2012-03-21 01:34 & 0.43 & 0.17 & 31 & 1:1.29 & 31 & Y \\

11446 & N31 E20 & 2012-03-22 15:58 & 2012-03-24 15:58 & 2012-03-25 11:10 & 2012-03-26 03:10 & 1.50 & 0.24 & 24 & 1:2.26 & 6 & Y \\

11480 & S14 E26 & 2012-05-09 11:10 & 2012-05-11 04:46 & 2012-05-11 11:10 & 2012-05-13 23:58 & 0.29 & 0.17 & 25 & 1:2.38 & 40 & N \\

11561 & S18 E34 & 2012-08-29 19:10 & 2012-08-31 04:46 & 2012-08-31 04:46 & 2012-09-01 14:22 & 0.99 & 0.33 & 28 & 1:2.55 & 36 & Y \\

11680 & S25 E52 & 2013-02-24 14:22 & 2013-02-25 14:22 & 2013-02-25 15:58 & 2013-03-03 03:10 & 0.82 & 1.08 & 50 & 1:0.03 & 28 & Y \\

11808 & N12 E66 & 2013-07-29 01:34 & 2013-07-30 01:34 & 2013-07-30 19:10 & 2013-08-01 11:10 & 2.13 & 0.85 & 29 & 1:2.57 & 18 & N \\

11813 & S19 E22 & 2013-08-06 01:34 & 2013-08-08 17:34 & 2013-08-08 17:34 & 2013-08-11 23:58 & 1.05 & 0.83 & 31 & 1:1.19 & 37 & Y \\

11867 & N17 E05 & 2013-10-11 07:58 & 2013-10-13 15:58 & - & - & - & - & - & - & - & - \\

11881 & S25 E52 & 2013-10-24 01:34 & 2013-10-26 15:58 & 2013-10-26 15:58 & 2013-10-28 14:22 & 1.04 & 0.49 & 34 & 1:0.94 & 25 & N \\

11886 & N10 E14 & 2013-10-28 09:34 & 2013-10-30 17:34 & 2013-10-30 17:34 & 2013-11-01 04:46 & 0.15 & 0.54 & 24 & 1:2.26 & 23 & Y \\

12086 & N03 E49 & 2014-06-08 15:58 & 2014-06-09 23:58 & 2014-06-10 15:58 & 2014-06-11 15:58 & 1.00 & 0.24 & 30 & 1:1.67 & 7 & Y \\

12119 & S26 E38 & 2014-07-18 04:46 & 2014-07-21 06:22 & 2014-07-21 06:22 & 2014-07-23 01:34 & 0.66 & 0.28 & 11 & 1:7.05 & 1 & Y \\

12168 & N10 E08 & 2014-09-16 12:46 & 2014-09-18 11:10 & 2014-09-19 20:46 & 2014-09-22 14:22 & 0.87 & 0.77 & 37 & 1:0.79 & 37 & Y \\

12229 & S23 E50 & 2014-12-04 20:46 & 2014-12-05 22:22 & 2014-12-06 04:46 & 2014-12-07 09:34 & 0.83 & 0.24 & 32 & 1:1.3 & 34 & Y \\

12273 & N02 E21 & 2015-01-15 15:58 & 2015-01-27 19:10 & 2015-01-27 19:10 & 2015-01-29 03:10 & 0.88 & 0.28 & 10 & 1:7.74 & 34 & Y \\

12274 & N03 E09 & 2015-01-25 17:34 & 2015-01-26 15:58 & 2015-01-26 15:58 & 2015-01-15 01:34 & 0.34 & 0.12 & 28 & 1:1.41 & 23 & N \\

12336 & N17 E49 & 2015-05-01 14:22 & 2015-05-05 20:46 & 2015-05-05 20:46 & 2015-05-08 23:58 & 0.70 & 0.53 & 18 & 1:3.42 & 1 & Y \\

12382 & S08 E29 & 2015-07-04 03:10 & 2015-07-05 20:46 & 2015-07-05 20:46 & 2015-07-09 01:34 & 3.50 & 0.27 & 42 & 1:0.32 & 19 & N \\

12453 & N04 E29 & 2015-11-12 07:58 & 2015-11-15 15:58 & 2015-11-15 15:58 & 2015-11-16 23:58 & 1.58 & 0.51 & 30 & 1:1.35 & 32 & Y \\

12455 & N14 E61 & 2015-11-13 04:46 & 2015-11-16 03:10 & 2015-11-16 06:22 & 2015-11-18 01:34 & 1.16 & 0.50 & 35 & 1:0.84 & 54 & Y \\

\hline
\end{tabular}
\caption{Flux cancellation of the 20 ARs in this study. The table shows the active region number as issued by NOAA, the location of the centre of the active region at the time of emergence, the approximate time of active region emergence, peak unsigned flux, start and end of the flux cancellation measurement. This is followed by the flux cancellation values including flux cancellation rate, total flux cancelled, percentage of the peak active region flux cancelled and the ratio of the flux cancelled vs. the flux remaining in the overlying arcade. The flux cancellation values are calculated in the time period between the start and end times of flux cancellation given in columns five and six. Also given is the absolute value of the shear angle of the active region at the time of peak flux, and whether convergence of the active region bipole is observed. \label{tab:table3}}
\end{rotatetable*}
\end{table*}

\begin{table*}[ht]
\centering
\begin{tabular}{|c|c|c|c|c|c|}
\hline
NOAA & Flux Cancellation & Total Flux & Total Percentage of & FR vs. Overlying \\
AR &  Rate (10$^{19}$~Mx~h$^{-1}$) & Cancelled (10$^{21}$~Mx) & peak AR flux (\%) &  Arcade Ratio \\
\hline
11437 & 0.43 & 0.17 & 31 & 1:1.29 \\
11561* & 0.99 & 0.42 & 28 & 1:1.57 \\
11680 & 0.82 & 1.08 & 49 & 1:0.03 \\
12382 & 0.35 & 0.27 & 43 & 1:0.32 \\
\hline
11446 & 1.50 & 0.24 & 23 & 1:2.26 \\
11808 & 2.13 & 0.85 & 22 & 1:2.57 \\
11881 & 1.04 & 0.49 & 34 & 1:0.94 \\
11886 & 0.15 & 0.54 & 24 & 1:2.26 \\
12336 & 0.70 & 0.53 & 18 & 1:3.42 \\
\hline

\end{tabular}
\caption{\label{tab:flux_canc} Flux cancellation values for the active regions that produced low-altitude eruptions originating from the internal PIL (top section) and the active regions that produced high-altitude eruptions (bottom section). The table shows the flux cancellation rate, total amount of flux cancelled, total percentage of peak unsigned active region flux cancelled and the ratio of the flux available to be built into the flux rope compared to the flux contained in the overlying arcade, i.e. the flux of the remaining bipole. *Corrected values for total flux cancelled, total percentage of peak active region flux and the ratio of the flux in the flux rope vs. overlying arcade are given for AR 11561.}
\end{table*}

\acknowledgments
The authors are grateful to the SDO/HMI and AIA consortia for the data, and also JHelioviewer (http://jhelioviewer.org/), which was used for browsing data. This research also makes use of SunPy \citep{SunPy-2015} (http://sunpy.org/), an open-source and free community-developed solar data analysis package written in Python. We are also grateful to Sally Dacie who helped with code development. S.L.Y. acknowledges STFC for support via PhD studentship and the Consolidated Grant SMC1 YST025. S.L.Y and L.M.G. are part of an International Space Science Institute (ISSI) team entitled ``Decoding the Pre-Eruptive Magnetic Configurations of Coronal Mass Ejections". We are thankful to ISSI for hosting us, Angelos Vourlidas and Spiros Patsourakos for leading the team, and the rest of the group for valuable discussions. L.M.G. acknowledges support through a Royal Society University Research Fellowship. L.v.D.G. is partially funded under STFC consolidated grant number ST/N000722/1. L.v.D.G. also acknowledges the Hungarian Research grant OTKA K-109276. L.M.G. and L.v.D.G. acknowledge funding under Leverhulme Trust Research Project Grant 2014-051. D.H.M. would like to thank both the STFC and Levehulme trust for financial support.


\bibliographystyle{yahapj}
\bibliography{references}

\end{document}